\documentclass[preprint,pre,amsmath,amssymb,showpacs]{revtex4}

\usepackage{graphicx}
\usepackage{dcolumn}
\usepackage{bm}

\def\kT{{k_{{}_B}T}}
\def\gsim{\mathrel{\raise.3ex\hbox{$>$\kern-.75em\lower1ex\hbox{$\sim$}}}}
\def\lsim{\mathrel{\raise.3ex\hbox{$<$\kern-.75em\lower1ex\hbox{$\sim$}}}}
\def\L{{\mathcal L}}
\def\G{{\mathcal G}}
\def\eq#1{{Eq~(\ref{#1})}}
\def\xhat{{\mathaccent 94 x}}
\def\yhat{{\mathaccent 94 y}}
\def\r{{\bf r}}

\begin{document}

\title{Interactions between proteins bound to biomembranes}
\author{A.R. Evans${}^1$, M.S. Turner${}^1$ and P. Sens${}^2$}
\affiliation{1 Department of Physics, University of Warwick, Coventry CV4
7AL, UK}
\affiliation{2 Institut Charles Sadron, 6 rue Boussingault, 67083
Strasbourg Cedex, France}

\date{December 23, 2002}



\begin{abstract}
We study a physical model for the interaction between general inclusions
bound to fluid membranes that possess finite tension $\gamma$, as well as the
usual bending rigidity $\kappa$. We are motivated by an interest in proteins
bound to cell membranes that apply forces to these membranes, due to either
entropic or direct chemical interactions. We find an exact analytic
solution for the repulsive interaction between two similar circularly symmetric
inclusions. This repulsion extends over length scales
$\sim\sqrt{\kappa/\gamma}$ and
contrasts with the membrane-mediated contact attraction for similar
inclusions on
tensionless membranes. For non circularly symmetric inclusions we study the
small,
algebraically long-ranged, attractive contribution to the force that
arises. We discuss the
relevance of our results to biological phenomena, such as the budding of
caveolae from cell
membranes and the striations that are observed on their coats. These, and
other, ``gnarly
buds" may prove fascinating to study further.
\end{abstract}

\maketitle

\section{Introduction}\label{introduction}

A significant proportion of all proteins in a typical eukaryotic cell are
membrane proteins. These are found anchored to cell membranes. Many of these
carry out tasks such as signal transduction, pore or channel formation,
cytoskeletal binding etc\cite{alberts}. Others are involved in endocytosis and
exocytosis. In particular it is now understood that the formation of clathrin
coated pits is driven by the controlled geometric aggregation of clathrin
which exert corresponding forces on the cell
membrane\cite{clathrin1,clathrin2}.
There also
exists a less well understood class of membrane invaginations, known as
caveolae, that are less morphologically distinct than clathrin coated
pits and resemble aspherical invaginations with a typical size of the order of
100nm\cite{rothberg}. There is good evidence that caveolae are involved in
endocytosis\cite{function1} and play an important role in cell
signaling\cite{review1}. Intriguingly they are probably also involved in the
sensing of shear-stress\cite{shear,shear2}. Such stresses would be expected to
act on the tension $\gamma$ of the cell membrane. Thus, the present work may
yield predictions for some of the direct physical consequences of surface
tension, e.g.
on inter-protein forces. In
section~\ref{analysis} we discuss some qualitative effects of surface tension
although a quantitative analysis of the effect of shear on membrane tension
is beyond
the scope of the present work. Recent elegant experiments demonstrate that
better control
of the surface tension may be achieved via micropipette aspiration
facilitating studies
of the effects of tension on membrane
elasticity\cite{rawicz,albersdorfer,hackl} and
permeability to water \cite{olbrich}.

It has now been shown that caveolins\cite{lisantiJCB94}, a recently discovered
class of membrane-bound proteins, are necessary for the formation of caveolae.
These caveolins have a short membrane spanning sequence and N- and
C-terminus polypeptide ``tails", totalling about 150 amino acids, {\em both}
found on the cytoplasmic side of the cell membrane. It is thought that these
caveolin molecules are typically found in small aggregates of size $b$ (a few
nm) containing of approximately 15 molecules\cite{lisantiPNAS95}. A crude
thermodynamic (entropic) bound on the bending moment would be $\gsim \kT b$,
higher for larger aggregates.

Many theoretical studies have sought to calculate the effect of adsorbed
polymer on flexible, fluid
membranes, see e.g. \cite{lipowsky2,carlos,lipowsky1a,lipowsky1b,kim}. One
reason for
this approach is that the  simplified models available from polymer physics
allow for a
more or less exact computation of the entropic pressure exerted by the
polymer on the
membrane. Our work represents a generalization of this approach. An
anchored polymer is
but one example, or model, of a membrane inclusions that exerts a force on
the membrane.
Any membrane protein that isn't perfectly symetrical across the membrane should
exert forces, the distribution of which will be dictated by the precise
protein configuration. Furthermore, it can be argued that the physical effect
of integral membrane protein can also be described by a well chosen
force distribution. Hence there is a need for a general theory on the
effect of arbitrary force distributions on flexible membrane. Our premier interest are large  peripheral membrane protein such as caveolin. Our theory is, however, general enough to be applied to any flexible fluid interface with fixed tension, containing embedded impurities.

Much of the previous work on interaction between membrane-bound objects has
focused on
membranes with vanishing surface
tension\cite{lipowsky2,carlos,lipowsky1a,lipowsky1b}, although there have been
exceptions \cite{golestanian}. This work exists within the context of
an extensive literature on membrane mediated interactions on tensionless
membranes, see e.g. \cite{goulian,goulianepl,goulianeplerratum}, which
focussed mainly on
Casimir-like forces originating from membrane fluctuations. It was argued
that these
could be either attractive or repulsive depending on the temperature and
certain details
of the model. The interest in tensionless membranes has its origin in the
fact that
isolated, self-assembled lipid membranes should be tensionless at equilibrium.
Importantly this is not the case for many cell membranes. Other studies
include tension
as a (shape dependent) parameter determined self-consistently for closed
membrane
surfaces\cite{seifert}. A study of the behavior of membranes under constant
surface
tension may shed light on the physics occurring on biological membrane,
which are not
truly at equilibrium and hence bear substantial tensions\cite{sheetz,lorz}.
Cells
commonly adjust their surface tension to a set value via a mechanism known as
Surface-Area-Regulation\cite{SAR}. Hence membrane phenomena over
sufficiently long
timescales effectively occur at constant surface  tension. Previous studies
of the
effects of rod-like objects embedded in fluid membranes\cite{golestanian}
and impurities
at gas-liquid interfaces\cite{fournier,stamou} share certain similarities
with this
study.

\section{Analysis of membrane deformation and membrane mediated
interactions}\label{analysis}

Our aim is to construct a theoretical model for the membrane-mediated
interactions between proteins bound to cell membranes. To achieve this we will
denote the normal membrane displacement from its average flat state as
$u(\r)$, where $\r$ is a 2D vector in the plane of the unperturbed membrane.
Motivated by a desire to analyze small membrane deviations we employ an
expansion of the free energy of the membrane in powers and gradients of
$u$. The free energy of a deformed fluid membrane, without inclusions, is well
approximated by\cite{safran}
\begin{equation}
F_{memb}=\int d^2r \left[\frac\kappa{2}(\nabla^2 u(\r))^2
+\frac\gamma{2}(\nabla u(\r))^2\right]\label{fmemb}
\end{equation}
Which represents the truncation of an arbitrary expansion at order $u^2$ (odd
orders are excluded by symmetry) and at the second derivatives of $u$. For a
planar membrane, with normal in the $\mathaccent 94 z$-direction, the gradient
operator is
$\nabla=\xhat \frac\partial{\partial x}+\yhat \frac\partial{\partial y}$.
The only two parameters that are needed to describe the physics of the
membrane are
$\kappa$, the elastic bending modulus (typically\cite{evans} $20\kT$ for
biomembranes) and $\gamma$ the surface tension (estimated to be \cite{sheetz}
$10^{-1}$ to $10^{-2}$pN/nm).

To include the work done by the pressure (normal force per unit area)
$f(\r)$ exerted by one or more inclusions we exploit the
fact that, for small deviations, the work done is merely the surface integral
of the product of pressure and distance. The displacement has no effect on the
pressure field at this order. Thus the total free energy, including the
effect of inclusions is
\begin{equation}
F=\int d^2r \left[\frac\kappa{2}(\nabla^2 u(\r))^2
+\frac\gamma{2}(\nabla u(\r))^2 - f(\r)u(\r)\right]
\label{f}
\end{equation}
where the term involving $f(\r)$ is chosen to have a
minus sign to ensure that the membrane displacement has the desired sign. At
thermodynamic equilibrium the free energy $F$ must reach its minimum value, at
which point $u(\r)$ satisfies the Euler-Lagrange equation
\begin{equation}
\L_1 \L_2 u(\r) - \frac{f(\r)}{\kappa} = 0 \label{euler}
\end{equation}
with $\L_1=\nabla^{2}$, $\L_2 = \nabla^{2} -
k^{2}$ and $k^2=\gamma/\kappa$ indicating an intrinsic length scale
$k^{-1}\approx 30$nm (with perhaps $6{\rm nm}\lsim k^{-1}\lsim 100{\rm
nm}$). We solve
this under the boundary condition
$\lim_{\r\rightarrow\infty} \nabla u(\r) = 0$ and
that $\lim_{\r \rightarrow 0}u(\r) $
exists. Since the commutator $\left[\L_1,\L_2 \right] = 0$ it can be shown that
\begin{equation}
u(\r) = \frac{1}{\gamma}\int\left[ G_{1}(\r,\r')-
G_{2}(\r,\r')\right]f(\r')d^2\r',
\end{equation}
where $G_1$ and $G_2$ are the Green's functions
corresponding to the self adjoint problems determined
by $\L_1$ and $\L_2$ (and their boundary conditions)
respectively. We now seek to find the Green's functions of these
self adjoint operators.

For $\L_1$ we solve
\begin{equation}
\L_1 G_{1}(\r,\r') = \nabla^2 G_{1}(\r,\r') = -\delta(\r - \r')\label{poisson}
\end{equation}
The only boundary conditions compatible with this
equation are Neumann conditions at infinity i.e.
$\lim_{{\r-\r'}\rightarrow\infty}\nabla
G_{1}(\r-\r') = 0$. This corresponds to the membrane being asymptotically
flat and yields
the well known Green's function of Poisson's equation in 2-D
\begin{equation}
G_{1}(\r,\r') = -\frac{1}{2\pi}\log{A|\r - \r'|}\label{Gone}
\end{equation}
where $A$ is an arbitrary constant with dimensions
of (length)$^{-1}$. We choose $A = k$  for
convenience, this choice corresponds only to a
definition of the zero of the free energy.

For $\L_2$ we must solve
\begin{equation}
\left(\nabla^2 - k^2\right)G_{2}(\r,\r') = -\delta(\r - \r').
\end{equation}
subject to the boundary conditions specified above.
This self adjoint problem has linearly
independent solutions $K_{0}(k|\r-\r'|)$ and
$I_{0}(k|\r-\r'|)$. These are, respectively,
the zeroth order modified Bessel functions of the first
and second kind. The function $I_{0}(x)$ diverges as
$x\rightarrow\infty$ therefore we write the
Green's function as
\begin{equation}
G_{2}(\r - \r') = \frac{1}{2\pi}K_{0}(k|\r-\r'|)
\end{equation}
This gives us the solution of  equation (\ref{euler}), subject
to the stated boundary conditions as
\begin{equation}
u(\r) = \int G(\r,\r') f(\r') d^2\r'\label{usolution}
\end{equation}
with the Greens function
\begin{equation}
G(\r,\r') = -\frac{1}{2\pi
\gamma}\left[ K_{0}(k|\r -\r'|) + \log{k|\r-\r'|} \right]\label{ugreens}
\end{equation}

Equations ~\ref{usolution} and \ref{ugreens} represent a solution for the
equilibrium membrane displacement $u(\r)$ due to an arbitrary force
distribution. The formal analysis is concluded in the appendix, where
we show that this solution is unique.

\subsection*{Interaction Potential between two proteins}

Consider a fluid membrane with two, not necessarily identical, inclusions bound
to it. Denote the pressure fields acting on the membrane surface due to each
inclusion by $\psi^{(1)}$ and $\psi^{(2)}$. Without loss of generality we may
let $\psi^{(1)}$ be centered at the origin and let the vector $\r$ joining the
centers of the pressure distributions lie along the x-axis. Mathematically this
means that we write the total pressure distribution as

\begin{equation}
f(\r') = \psi^{(1)}(\r') + \psi^{(2)}(\r' -\r),
\end{equation}
where $\psi^{(2)}(\r'-\r)$ is the pressure distribution of the second inclusion
centered at $\r$, rather that the origin. Hence the Free Energy at
equilibrium is
given by $F= F_{self}^{(1)} +F_{self}^{(2)} + \Phi(\r)$, where
$F_{self}^{(i)}$ is the self energy of the i$^{{\rm th}}$
inclusion and $\Phi(\r)$ is the interaction energy. We are
interested only in $\Phi(\r)$ as this is what determines the
physics of interest to us here. By direct analysis or analogy with
electrostatics this may be expressed as

\begin{equation}
\Phi(\r) = -\int d^2\r' \int d^2\r''
\psi^{(1)}(\r') \psi^{(2)}(\r'') G(\r -\r' +\r'')\label{phidef}
\end{equation}
where $G(\r)$ is the real space Green's function.

For the case that we have two inclusions with circular symmetry
we are
able to determine the interaction potential exactly. Consider two circularly
symmetric inclusions i.e.
$\psi^{(i)}(\r')=\psi^{(i)}(|\r'|)=\psi^{(i)}(r')$
such that $\psi^{(i)}(r')=0$ for $r'>b$, where $b$ is some maximum radius
of the
inclusion. Also each pressure field is such that no average force
acts ($\int\psi^{(i)}(\r')d^{2}\r'=0$). For inclusions which are not
anchored to external structures, such as the
cytoskeleton, this condition must be satisfied on general grounds. Indeed
Newton's third
law requires that any average force applied to the membrane by an inclusion
must be
equal and opposite to the reaction force applied by the inclusion to some
external
structure. If there is no external structure for this reaction force to act
against then
there can be no average force on the membrane. We assume that the pressure
fields do not
overlap so that
$r>2b$. Under these assumptions the logarithmic term in
\eq{ugreens} gives a vanishing contribution to the integral \eq{phidef} due
to Gauss'
Theorem. It corresponds to the electric field due to two circularly
symmetric charged
discs in 2-D, each carrying no average charge and thus with zero field in
the region
outside both discs. This analogy will be revisited when
we discuss non-circularly symmetric inclusions in
section~\ref{multipole} below. The potential between two
circularly symmetric inclusions is therefore given by
\begin{equation}
\Phi(\r) =
\frac{1}{2\pi\gamma}\int_{0}^{b}r'dr'\int_{0}^{b}r''
dr''\psi^{(1)}(r')\psi^{(2)}(r'')\int_{0}^{2\pi}d\theta '
\int_{0}^{2\pi}d\theta '' K_{0}(k|\r-(\r'-\r'')|),
\end{equation}
the upper limits of integration are both $b$ since
$\psi(r')=0$ for $r>b$. We make use of the identities
\begin{equation}
\int_{0}^{2\pi}K_{0}(k|\r-(\r'-\r'')|)d\theta '' =
2\pi K_{0}(kr) I_{0}(k|\r'-\r''|)
\end{equation}
valid for $r>|\r' -\r''|$, and
\begin{equation}
\int_{0}^{2\pi}I_{0}(k|\r'-\r''|)d\theta ' =
2\pi I_{0}(kr') I_{0}(kr'')
\end{equation}
valid for $r'\ne r''$. Thus we are able to derive an exact analytical
form for the interaction potential between non-overlapping ($r>2b$) circularly
symmetric inclusions
\begin{equation}
\Phi(\r) = \frac{1}{2\pi\gamma} \zeta^{(1)}\zeta^{(2)} K_{0}(kr)\label{phisymm}
\end{equation}
where $\zeta^{(i)}= 2\pi\int_0^\infty r'\psi^{(i)}(r')I_0(kr')dr'$
characterizes the strength of the i${}^{\rm th}$ membrane/inclusion
coupling. By employing the expansion of $I_0(kr')$ the quantity
$\zeta^{(i)}$ can
be thought of as a series of moments of the force distribution. The radial
force
is, as usual given by the derivative of the potential
$f_r=-{\partial\Phi}/{\partial r}$. The interaction is everywhere {\it
repulsive}.
Some previous  studies have reported attractive interactions between
polymer chains
grafted on tensionless membranes\cite{lipowsky2,carlos}. In these studies
the position
$u(0)$ of the chain grafting point was fixed, effectively by a Lagrange force
which then acted to ensure that there was no average force, although this
condition was
not identified explicitly by these authors. In the case of vanishing
tension it was
found that there was no interaction between inclusions unless they
``touch", i.e. their
force fields overlap. One can understand this as follows: For tensionless
membranes the
membrane deformation under the force field is dictated by the minimisation
of an energy
similar to \eq{f} but with
$\gamma=0$. The membrane deformation outside the extent of the force field
simply
minimises the curvature energy, hence has a zero mean curvature and does
not  contribute
to the total energy. As two inclusions approaches each other, the total
energy does not
change until one force field couples to the membrane deformation directly
under the
other force field,  since the deformation energy of any piece of membrane
that it not in
direct contact with a force field is zero. There is no characteristic
lengthscale
$k^{-1}$ to give the interactions finite range when $\gamma=0$. In consequence,
inclusions  interact only when the two force fields overlap. in the case of
grafted
ideal polymer chains, the energy is decreased when the force  fields
overlap because
they both want to deform the membrane in the  same direction, resulting in an
attraction. We note that the interaction between overlapping membrane
inclusions is very sensitive  to the direct physical, rather than membrane
mediated,
interactions between them, unlike
\eq{phisymm} which is universal. In the
$\gamma\to 0$ limit
\eq{phisymm} gives vanishing interactions, entirely consitant with these
earlier results
for tensionless membranes.

As mentioned in the introduction recent experiments on the effect of shear
forces on living cells provide a crude control on membrane tension.
In future studies we plan to calculate the precise
effect of this on, e.g. surface phase equilibria and budding phenomena.

\section{Microphase separation at the surface of caveolar
invagination}\label{stripesection}
Membrane-mediated interactions between membrane-bound inclusions have
been extensively studied theoretically \cite{goulian} with, however,
relatively little
discussion of how experiments could easily support these predictions.
Although presumably long range, these interactions are expected to be quite
small, and
are probably often dominated by stronger, (bio)chemically specific, short range
phenomena. Long range interactions can, however, profoundly influence the
phase
behaviour of certain membrane proteins even if stronger, short range
interactions, are present. As we emphasise below studies of such surface
phases might
provide an indirect probe of the membrane mediated interactions of interest
here. This
might be of particular relevance for  the study of the phase behaviour of
the protein
caveolin, which is  found at high density on certain membrane invaginations
called
caveolae (see sec.I).

One peculiar feature of the caveolae bulbs is
their texture. Distinct striations are seen at
the surface of these buds\cite{rothberg} (see also Fig.13.48 in
\cite{alberts}). These are
now thought to correspond with the organization and alignment of caveolin
oligomers on the
membrane \cite{cterm1b}. The observation of these surface stripes is
intruiging, and the reader may find it interesting that radially
symmetrical oligomers can give rise to non-symmetrical surface phases. We
discuss this below, arguing  that
the stripe phase
may be a signature of the membrane mediated repulsion between protein
aggregates, such as
those calculated in section~\ref{analysis}. Indeed, studying the phase
behaviour of membrane inclusions may provide one of the best ways to test
theory
against experiment in this field. Molecular dissection of the caveolin
protein has shown
that caveolin oligomers strongly attract each other through contacts of
specific protein sequences\cite{review1} (a short range attraction). It is
well known that a
solution of particle interacting via hard core repulsion and a short
range attraction undergo a gas/liquid phase separation.
This results in
large dense  regions (the liquid) coexisting with less dense regions (the
gas). The
caveolin oligomer being anchored to the cell membrane, there exists an
additional
membrane-mediated, longer range repulsion between between them, as
demonstrated in
section~\ref{analysis}, which allow for a more complex phase
behavior\footnote{A number
of other factors, including variation in the membrane composition, may
influence the phase
behavior of these objects}. It has been recently argued at the light of
computer
simulations\cite{gelbart,searPRE,searJCP} that a long range repulsion
should break the gas
and liquid phases into microdomains. The short range attraction still
locally drives a
gas/liquid phase separation, but large aggregates are costly because of the
long range
repulsion. The microdomains are circular liquid islands at low density,
circular gas
regions at high density, and stripes otherwise.

To gain a quantitative understanding of how stripes may form on caveolae we
study the
stability of an homogenous (2-D) solution of oligomers of area
$s_b=\pi b^2$ interacting via a potential $V(\r)$, which include a
short range (the particle size $b$)
attractive exponential interactions  and the longer range membrane
mediated repulsion of \eq{phisymm}: $V(r)=-E_a e^{-r/b}+E_r
K_0(kr)$, where $E_a$ and $E_r$ are respectively the strength of the
attraction and the repulsion. We look at small perturbations
$\delta\phi$ (with the conservation
rule $\int
dS\delta\phi(r)=0$) around the average surface coverage $\phi_0$. The
free energy includes the pair interaction and  the translation
entropy of the inclusions, for which we use the gas-on-a-lattice
model. Expanding the free energy:
\begin{eqnarray}
F=\int
\frac{dS}{s_b}\left(\phi\log{\phi}+(1-\phi)\log{(1-\phi)}\right)+\frac12\int\frac{dSdS'}{s_b^2}\phi(r)
V(|r-r'|)\phi(r')\cr
\delta
F=\int\frac{dS}{s_b}\frac{\delta\phi^2}{2\phi_0(1-\phi_0)}+\frac12\int\frac{dSdS
'}{s_b^2}\delta\phi(r)
V(|r-r'|)\delta\phi(r')
\end{eqnarray}
The Fourier transform $\delta\phi(r)=S\int
\frac{d^2q}{(2\pi)^2}\delta\phi_q e^{i{\bf
qr}}$ allows us to investigate the formation of structures. It leads to
\begin{equation}
\delta F=\frac12\frac{S^2}{s_b}\int\frac{d^2q}{(2\pi)^2}{\cal
V}_q|\delta\phi_q|^2\qquad {\rm with}\qquad {\cal
V}_q=\frac{T}{\phi_0(1-\phi_0)}+\frac1{s_b}V_q
\end{equation}
If $V_q$, the Fourier transform of $V(\r)$, is sufficiently negative
(attractive) for a
given mode
$q$ then ${\cal V}_q<0$ and the mode is unstable. Substituting the
expression for $V(\r)$, we obtain
\begin{equation}
{\cal V}_q=2k_BT\left[\frac{1}{2\phi_0(1-\phi_0)}+\left(-E_a\frac{1}{(1+(b
q)^2)^{3/2}}+E_r\frac{1}{k^2+q^2}\right)\right]
\label{stripes1}
\end{equation}

In the absence of long range repulsion, the most unstable mode
is always $q=0$ (macrophase separation), and the liquid/gas
transition is observed provided $2E_a\phi_0(1-\phi_0)>\kT$.
Because
of the membrane-mediated long range repulsion, the function ${\cal V}_q$
presents a
minimum if $E_r>\frac{3}{2}E_a(kb)^4\sim 10^{-3} E_a$ (for the
typical numbers $b\simeq 5$nm and $k^{-1}\sim 30$nm). One sees that
although the strength of
membrane-mediated interaction (\eq{phisymm}) is expected to be quite
small:  $E_r\simeq 10^{-2}\kT$ (see sec. V.A, or ref.\cite{usPNAS}
for a more detailed analysis), it is of much longer range, and should
be competent to
produce a well ordered phase.

The phase separation occurs
preferentially for a mode given by $\partial{\cal V}_q/\partial q=0$,
and periodic
arrays of dense and dilute regions are observed.  In the limit
$kb\ll1$ the structure have a typical size
$2\pi/q^*$ which is independent of the range of the repulsion:
$q^*b=(2E_r/3E_a)^{1/4}\simeq \frac{1}{4}$. It thus defines dense
stripes of width about $5$ particle diameters, which agrees
quite well with the experimental observations. The existence of a
stripe phase at the surface of the biological structures known as
caveolae is thus quantitatively consistent with the existence of a
membrane-mediated repulsion between proteins described by
\eq{phisymm}.

This analysis does not help us to discriminate between circular domains and
stripes. Computer simulation have clearly shown that this transition indeed
exists for high
enough surface coverage\cite{searPRE,searJCP}.

Note that the formation of striped mesostructures in membrane have been
predicted on the
basis of  curvature effects only\cite{leibler,leiblerandelman}. However, it
can be shown
that their existence at the surface of invagination of such large curvature
as the caveolae
($R\sim 50$nm) is inconsistent with the existence of invaginations with a
well defined size.

\section{Far field interactions of non circularly symmetric
inclusions}\label{multipole}

If the force distribution of the inclusions is not circularly symmetric
then no general analytic solution to
\eq{phidef} exists. However we can still proceed by examining the
far-field interactions of slightly asymmetric inclusions. In  order to do this
we will parameterize the asymmetry of the $s^{\rm th}$ inclusion by
\begin{equation}
D^{(s)}_{ij}=\int d^2\r' r'_i r'_j
\psi^{(s)}(\r')=\left(
\begin{array}{cc}
c+\epsilon^{(s)}&0\\ 0&c-\epsilon^{(s)}
\end{array}
\right)\label{Dunrot}
\end{equation}
where $i$ and $j$ are cartesian indices, $c$ controls the average (isotropic)
magnitude and $\epsilon$ controls the anisotropy of a slightly ``elliptical"
force distribution, extended in the
$x$-direction and contracted in the $y$-direction (for $\epsilon>0$). This
force distribution can be rotated
by the usual rotation matrix
$R(\theta)$ to obtain
the effective moment $R_{im}(\theta^{(s)})R_{jn}(\theta^{(s)})D^{(s)}_{mn}$
of an
``elliptical" force distribution rotated by an angle $\theta^{(s)}$
relative to the $x$-axis. This
expression involves two rotations, one for each factor of $\r'$ in \eq{Dunrot}.

It is important to first note is that the interaction can be separated into two
terms, one that is an
integral involving $K_0(k|\r-\r'+\r''|)$ and the other that is an
integral involving  $\log k|\r-\r'+\r''|$. The first of these terms gives a
contribution to the interaction
potential that is dominated by a term like \eq{phisymm}, with corrections
due to the asymmetry that are
smaller (and also asymptotically exponentially short ranged). This conclusion
is independent of the precise
choice of asymmetry, provided only that it is small. The contribution from
the second term is more
interesting and can be shown to give rise to a small but {\it long ranged}
correction $\delta\Phi$ to the
interaction potential. This can be best understood by way of  multipole-like
expansion of $\log k|\r-\tilde\r|$ where $|\tilde\r|\ll|\r|$ and we choose
the direction of $\r=r\xhat$ to define the $x$-axis (the 1 direction), without
loss of generality. Thus $\tilde r_1$ and $\tilde r_2$ are the $x$ and $y$
components of $\tilde r$ and
\begin{equation}
\log k|\r-\tilde\r|=\log kr-{\tilde r_1\over r}+{\tilde r_2^2-\tilde
r_1^2\over 2 r^2}-{\tilde r_1^3-3\tilde
r_1\tilde r_2^2\over 3 r^3}+{6\tilde r_1^2\tilde r_2^2-\tilde
r_1^4-\tilde r_2^4\over 4 r^4}+O\left({1\over r^5}\right)\label{logexpansion}
\end{equation}
The conditions of no overall force $\int d^2\r' \psi(\r')=0$ and no overall
moment $\int d^2\r'\r' \psi(\r')=0$
for each inclusion mean that the first non-vanishing contribution to
$\delta\Phi\sim 1/r^4$, the
classical result for electrostatic quadrupole-quadrupole interactions in
two dimensions. Using
$\tilde\r=\r'-\r''$ in
\eq{logexpansion} to leading order in $1/r$ we have\footnote{From $\delta\Phi =
\frac{1}{4\pi\gamma r^4}\int d^2\r'\int d^2\r''
\psi^{(1)}(\r')\psi^{(2)}(\r'') {3\over 2}\left(4
x'x''y'y''-(x'^2-y'^2)(x''^2-y''^2)\right)$ hence $\delta\Phi
=\frac{1}{4\pi\gamma
r^4} R_{im}(\theta^{(1)})R_{jn}(\theta^{(1)})D_{mn}(\epsilon^{(1)})
R_{kp}(\theta^{(2)}) R_{lq}(\theta^{(2)})D_{pq}(\epsilon^{(2)}) \G_{ijkl}$
with the interactions defined by the kernal 
$\G_{ijkl}={3\over 2}
(4\delta_{i1}\delta_{j2}\delta_{k1}\delta_{l2}-(\delta_{i1}\delta_{j1}-\delta_{i2}\delta_{j2})(\delta_{k1}\delta_{l1}-\delta_{k2}\delta_{l2}))$ 
and the result follows by
contracting over all indices. }

\begin{eqnarray}
\nonumber\delta\Phi & = &\frac{1}{4\pi\gamma}\int d^2\r'\int d^2\r''
\psi^{(1)}(\r')\psi^{(2)}(\r'') \log k|\r-\r'+\r''|\\
& = &
\frac{-3\epsilon^{(1)}\epsilon^{(2)}}{2\pi\gamma
r^4}\cos2(\theta^{(1)}+\theta^{(2)})
\label{generalmultipolephi}
\end{eqnarray}
This result demonstrates that elliptical inclusions
attract if arranged so that
\mbox{$\theta^{(1)}+\theta^{(2)}=n\pi$} with $n$ an integer, as reported in
an earlier study of rod-like
inclusions\cite{golestanian}. This condition represents a degenerate family
of orientations in which the
orientation of the quadrupoles has reflectional symmetry about the mid plane.
This perturbative interaction is long-ranged, scaling like $1/r^4$.
Interestingly this dominates the exponentially short ranged repulsion from
\eq{phisymm} for large enough separations.

There is a straightforward way to understand the appearance of an algebraic
potential for anisotropic inclusions. The physics of \eq{poisson} and is
that of the Poisson equation in which the force
distribution is analogous to the electrostatic charge. The potential due to a
point charge in 2D is therefore \eq{Gone}. Gauss law tells us that there is no
field outside a circularly symmetric charge distribution that has no overall
charge (imbalance). This condition is analogous to our requirement here that
there be no overall force
$\int d^2\r' f(\r')=0$ by Newton's law. If there is no field the potential is
constant (zero without loss of generality)  and no forces act on the
inclusions. However, if we relax the condition that the force (charge)
distribution is circularly symmetric there is no longer a simple symmetry
argument that the radial and azimuthal components of the field must be zero.
Indeed they are not. As is usual in multipole expansions in electrostatics an
algebraic potential results. It is here analogous to the quadrupole-quadrupole
interaction in electrostatics, since there are no dipole moments $\int d^2\r'
\r'\>f(\r')=0$. This is due to the fact that there can be no external first
force
moment (torque) on the membrane if it is not anchored to any
external structure against which the moment can act.

The physical consequences of these interactions are potentially significant.
In spite of the fact that the dominant interaction is repulsive the surface
concentration of membrane inclusions may often be maintained at fairly high
surface fractions by the regulatory mechanisms of the living cell. The
composition of the surface coats of caveolae may be further enriched in
several important membrane components, including cholesterol and proteins in
the caveolin family, as well as others\cite{review1,lisantiJCB94}. This means
that there is an effective surface pressure, driven by a chemical potential
difference that forces the inclusions to partially overcome their repulsion.
In this situation the attractive
$O(\epsilon^{(1)}\epsilon^{(2)})$ interactions might become significant,
leading to an in plane anisotropic phase separation such as is observed in
model systems\cite{gelbart,searPRE,searJCP,seul} and on the coats of
caveolae\cite{rothberg}. For caveolae the typical interactions seem too
weak \cite{usPNAS} to be solely responsible for the stripe morphology,
which may rather be
dominated by specific attractions, see section~\ref{stripesection}
\cite{cterm1b}.
Finally the existence of an attractive interaction that can arise from
fluctuations in the
inclusion force distribution (shape), as parameterized by
$\epsilon$, suggests the possibility of an attractive,
fluctuation driven force reminiscent of Van der Waals forces. This
mechanism should extend to systems such as
wetting droplets at gas-liquid interfaces. These can potentially exert much
stronger forces on the
interface, resulting in a more exaggerated effect (see
section~\ref{wettingtheory}).

\section{Origin of the force distribution}

Up to this point we have merely postulated the existence of a force
distribution $\psi(\r')$ due to a membrane bound inclusion. This force can
arise from direct mechanical effects, such as due to the geometry and shape
of the inclusion, e.g. the protein clathrin, or entropic effects due to the
asymmetric
anchoring of, e.g. flexible hydrophilic polymers, to a membrane. The budding of
caveolae is now thought to be driven by the protein caveolin which is known
to form small
oligomers resembling a number of polypeptide chains extending only from the
cytoplasmic
side of the membrane\cite{lisantiPNAS95,monier95}.

We have in mind that we have demonstrated that interactions should exist in
general, but
haven't yet adressed the question of how big they might typically be. We are
particularly interested in any models that can be chosen to be broadly
comparable with
membrane proteins. In this section we will present two examples of models
for the origin
of a force distribution arising from anchored polymer chains. This approach
represents the application of a simple, and hence rather ``idealised",
theory drawn from
polymer physics. Its exact quantitative applicability to polypeptide chains
is almost
certainly limited to a small subset of peptide sequences which either
resemble a diffuse
random coil or a dynamic, dense hydrophobic globule. Nontheless in the
absence of any
detailed microscopic information on a specefic inclusion's configuration
and stability
the approaches that we outline below probably represent the best chance for
us to obtain
a rough idea of the likely scale of the effects that we have described in
this article.
It is quite possible that the forces exerted by membrane proteins with
significant
well defined tertiary structure may be somewhat higher than the estimates
that we will
present below. This encourages us to consider our estimates as approximate
lower bounds
on the scale that these effects might reach.

In polymer physics terminology the two models that we consider for the
force applied by
a membrane-bound polymer will correspond to the two limiting of ``good" and
``poor" solvent conditions respectively\cite{degennes}. We will discuss
below how these
models might apply to caveolin homo-oligomers. Other recent studies of the
physics of
flexible polymers grafted onto tensionless fluid
membranes\cite{lipowsky2,carlos,lipowsky1a,lipowsky1b} undoubtedly share
similar
motivation. One feature that these earlier studies all have in common is
that there is
only an attraction for polymer force distributions that overlap one-another
on the
surface of the membrane and, when this is the case, the interaction is {\it
attractive}.
This is opposite in sign to the interactions that we predict on membranes
that bear
tension, see \eq{phisymm}. Furthermore our interactions are extended, with
ranges $\sim
k^{-1}$.

\subsection{Anchored polymers and polymer aggregates: Good solvent
conditions}\label{brushtheory}

Motivated by possible biophysical relevance we investigate the effect of
flexible polymer chains anchored to a small patch of membrane, of radius $a$.
We assume that the chains are in a good solvent, i.e. that they are found in
extended, hydrated random coil configurations. The chains
form a hemisphere of outer radius $b>a$, see figure~1. This model may be
used to
treat polymer homo-oligomers made up of a general number $Q$ of chains. As such
it shares many features with those that are known for caveolin
homo-oligomers\cite{lisantiPNAS95}.

\begin{figure} 
\centerline{\includegraphics[width=15cm]{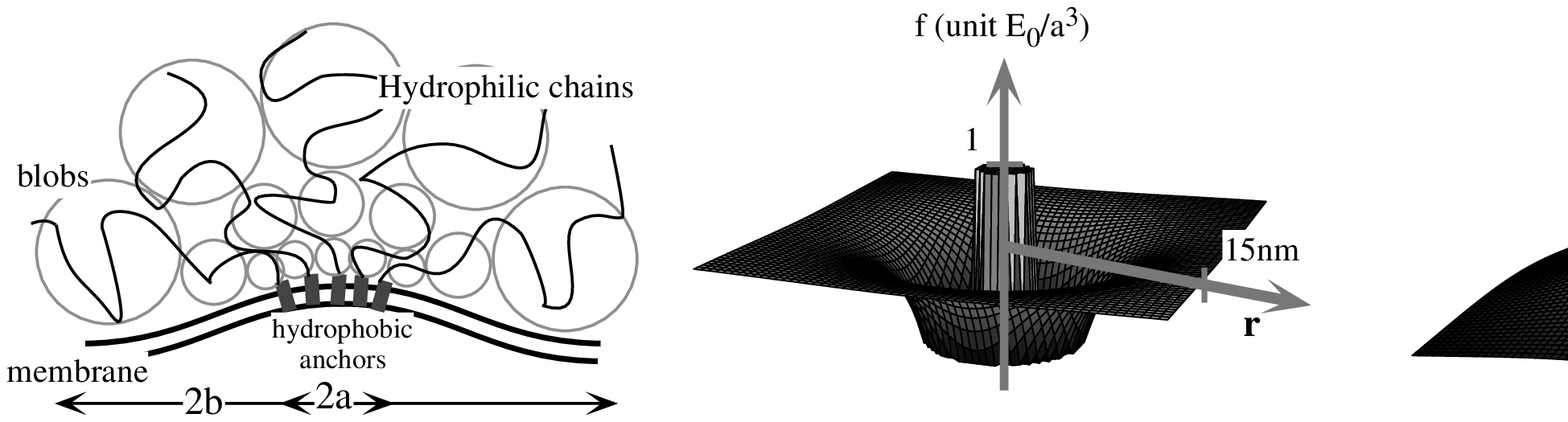}}
\vspace{2 cm}
\caption{Sketch of the form of the anchored polymer aggregate in good
solvent conditions,
when the polymer configurations are somewhat extended from the membrane.
Our model for the
force distribution $f(\r')$ is given by Eq~(\ref{cc}) and represents the
force per area
applied to the membrane by the polymers. The membrane is pushed down by the
``corona" of
the grafted polymers out to
$r=b=5$nm and is pulled upwards by the anchored ``core" inside $r'=a=2$nm
as a result.
For aggregates residing on the cytoplasmic face of the membrane, including
caveolin
homo-oligomers the cell interior would be above the membrane. The resulting
deformation
$u(\r')$ is shown out to $r'=15$nm for the
following values of the parameter values;
$\gamma=10^{-4}$J/m${}^{-2}$,
$\kappa=20\kT$ and
$Q=15$. The force and deformation can be expressed in terms of the
characteristic energy
$E_o$, see Eq~(\ref{cc}).}
\label{brush}
\end{figure}

Within the ``corona" of the polymer hemisphere, at radial distances
$b<r<a$ from the center of the polymer aggregate, there exists a characteristic
correlation length $\xi(r)$ for the polymer chains which crudely, represents
the spatial distance between collisions between the segments of a chain with
other chains or the membrane\cite{degennes}. The conservation of number of
chains implies\cite{daoudcotton} $\frac{1}{2}4\pi r^2 \simeq Q
\xi(r)^2$ for $a<r<b$. This is since the surface area of a hemisphere of radius
$r$ is filled by close packed blobs, up to a constant of order unity. From this
one may immediately deduce the scaling of the
correlation length $\xi(r) = \left(\frac{2\pi}{Q}\right)^\frac{1}{2} r$.
The work done in generating each blob is $k_{B}T$,
independent of the blob size. Thus we may write the
pressure in this region as the energy per blob
divided by the volume of a blob\cite{daoudcotton}
\begin{equation}
f(\r)=\frac{k_{B}T}{\xi(r)^3} =
\left(\frac{Q}{2\pi}\right)^\frac{3}{2}
\frac{k_{B}T}{r^3}
\end{equation}
for $a<r<b$. For $r>b$ the pressure is here assumed to be zero\footnote{In
nature there will often be a smooth crossover around $r=b$. Nonetheless we
consider our approximation to be as good as any alternative and insist that, at
large distances, the pressure must ultimately vanish due to the finite lengths
of the polymer chains}. The pressure in the core binding region is assumed
to be
constant and must involve a total force equal and opposite to that applied by
the corona so that no average force acts, as required by Newton's second law.
\begin{equation}
f(r) = {\Bigg\{}
\begin{array}{cl}
E_o a^{-3}& {\rm if}\quad 0<r<a\\
{-E_o\over{2(1-a/b)}}r^{-3}& {\rm if}\quad a<r<b\\
0 & {\rm if}\quad r>b
\end{array}
\label{cc}
\end{equation}
where $E_o=f(0)a^3=2k_B T(1-a/b)(Q/2\pi)^{3/2}$ is a characteristic energy
related to the
force applied at the center of the aggregate. This model represents one
approximate
framework for the understanding of forces applied by membrane biopolymers, see
Fig~\ref{brush}. It possesses several important features: It involves no
average force
being applied to the membrane. It involves no average first force moment by
symmetry.
The first non-vanishing moment is the second, corresponding to a bending
moment with
magnitude controlled by $\int r^2 f(r) d^2r\simeq E_o a$ with the
characteristic energy
scale $E_o\gsim\kT$ that is entropic in origin.

Models similar to this have been
proposed elsewhere\cite{lipowsky2,carlos,lipowsky1a,lipowsky1b} although,
as emphasized
previously, the physical behavior of these inclusions on fluid membranes that
are under tension is rather different.

We now proceed to give an estimate of the membrane
deformation and interaction potentials due to membrane-bound biopolymers
within the good
solvent polymer brush model described above.

For biological membranes primarily comprised of a bilayer of phospholipids the
tension and rigidites are typically\cite{evans,sheetz}
\begin{equation}
\kappa=10k_BT\ -\ 40 k_BT\quad\kappa_{bio}=20k_BT\qquad\gamma=10^{-5}\ -\
10^{-3}J/m^2\quad\gamma_{bio}=10^{-4}J/m^2
\end{equation}
where $x_{bio}$ refers to what we would take as a single ``typical" value
for biological
membranes. From this we obtain the lengthscale
\begin{equation}
k^{-1}=6nm\ -\ 100nm\quad  k^{-1}_{bio}=30nm
\end{equation}
which may crudely be identified with the range of the interactions.

\subsubsection{Membrane deformation due to the inclusion}

The deformation at $r=0$ is
\begin{equation}
u_{brush}(r=0)=\frac{E_0}{\kappa}\alpha
\end{equation}
in a frame where $u=0$ at infinity. This therefore represents the total
magnitude
of the normal deviation of the membrane expressed in terms of a
dimensionless ratio
$\frac{E_0}{\kappa}$ and a characteristic length $\alpha$.

For parameters that might be typical of caveolin homo-oligomers, $a=2{\rm
nm}$ and
$b=5{\rm nm}$, we find
\begin{equation}
\alpha(k^{-1}=6{\rm nm})=0.7{\rm nm}\quad\alpha(k^{-1}=30{\rm nm})=1.5{\rm
nm}\quad\alpha(k^{-1}=100{\rm nm})=2{\rm nm}
\end{equation}
much the same as for slightly larger inclusions with  $a=1nm$ and $b=10nm$
\begin{equation}
\alpha(k^{-1}=6{\rm nm})=1.3{\rm nm}\quad\alpha(k^{-1}=30{\rm nm})=2.7{\rm
nm}\quad\alpha(k^{-1}=100{\rm nm})=4{\rm nm}
\end{equation}

\subsubsection{Membrane mediated interaction energy between inclusions}

The interaction energy is expressed as
\begin{equation}
\Phi_{brush}=\frac{E_0^2}{\kappa}\beta
\end{equation}
which is the product of an energy ${E_0^2/\kappa}$, typically of the order of
$\kT$, and a dimensionless number $\beta$.
For $a=2{\rm nm}$ and $b=5{\rm nm}$
\begin{equation}
\beta(k^{-1}=6{\rm nm})=3\
10^{-2}\quad\beta(k^{-1}=30{\rm nm})=10^{-3}\quad\beta(k^{-1}=100{\rm
nm})=10^{-4}
\end{equation}
while for $a=1{\rm nm}$ and $b=10{\rm nm}$ we find instead
\begin{equation}
\beta(k^{-1}=6{\rm nm})=0.13\quad\beta(k^{-1}=30{\rm nm})=5\
10^{-3}\quad\beta(k^{-1}=100{\rm nm})=5\ 10^{-4}
\end{equation}

\subsection{Poor solvent conditions}\label{wettingtheory}

If the polymer chains are more hydrophobic, and the solvent conditions are
poor, the polymer chains may collapse into a tight, roughly hemispherical,
region
from which water is largely exuded. This represents the natural opposite limit
to the good solvent polymer model considered above. In the poor solvent case we
propose to model the inclusion as if it were a fluid droplet that partially
`wets' the membrane surface\cite{leger}. At any point on the circumference of
the drop the forces per unit length (due to the surface energy) are in
equilibrium. Resolving these force perpendicular to the membrane we find a
total force acting upward
\begin{equation}
F=2\pi\sigma\sin\theta
\end{equation}
where $b$ is the radius of the drop, $\sigma$
is the energy cost per unit area of producing the
interface between the inclusion and the external poor
solvent, typically water, and $\theta$ is the contact angle given by Young's
law\cite{leger}. Since we are in equilibrium this force must be
balanced by the force acting on the interface between the drop
(inclusion) and the membrane for $r<b$. To find the corresponding pressure
we need only consider the partial wetting droplet as if it formed part of a
larger sphere. The pressure is constant everywhere inside the sphere and is
given by the Laplace law
\begin{equation}
P=\frac{\partial E}{\partial V}=\frac{2\sigma}{R},
\end{equation}
where $R=b/\sin\theta$ is the radius of the sphere. The radius $R$ can
therefore be related to the volume of the drop (and $b$) by simple
geometry. Hence we may write the pressure distribution as
\begin{equation}
f(\r) = \frac{\sigma}{R}\left( b\>
\delta(r - b) - 2\Theta(b - r) \right)\label{dropforce}
\end{equation}
where $\Theta(x)$ is the Heaviside unit step function.

\begin{figure} 
\centerline{\includegraphics[width=15cm]{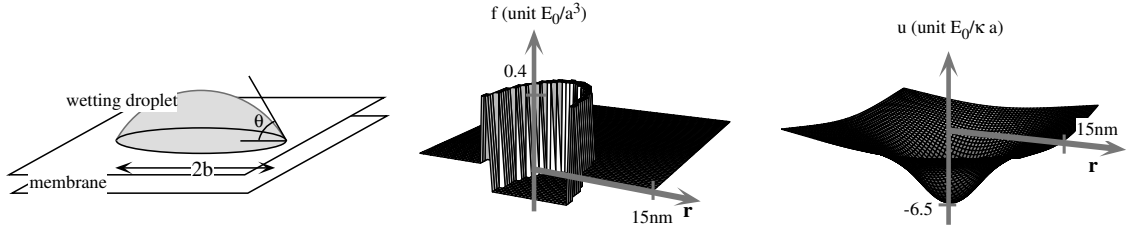}}

\vspace{2 cm}
\caption{Sketch of the shape of the anchored aggregate in poor solvent
conditions,
when the polymer forms a droplet on the membrane that is assumed to largely
exclude water.
Our model for the force distribution $f(\r')$ is given by
Eq~(\ref{dropforce}). The
membrane is now pulled upwards by the resolved Young's force on the contact
line at
$r'=b=5$nm and is pushed downwards by the Laplace hydrostatic pressure for
$r'<b$ as a result. The resulting deformation
$u(\r')$ is shown out to $r'=15$nm for the same total integrated force $\int
d^2r'|f(\r')|$ shown in Fig~\ref{brush}, for comparison. All the other
parameters values are
also as given in the caption to Fig~\ref{brush}. Even for the same
integrated absolute force the membrane deformation is an order of magnitude
larger than
shown in Fig~\ref{brush}, as well as having opposite sign. The
enhanced effect is due to the concentration of the forces at the exterior
of the aggregate.}
\label{drop}
\end{figure}

This model represents a different possible physical origin for the
pressure distribution that might be valid, e.g. for surface-anchored
hydrophobic polymers in
the poor solvent regime, when little water penetrates the protein chains, see
Fig~\ref{drop}. It also preserves the same features as the good solvent
polymer model: It
involves no average applied force or first force moment and has a bending
moment $\int r^2
f(r) d^2r\simeq E_o b$ that is enhanced by the localization of part of the
force
distribution around the exterior circumference $r=b$ of the inclusion's
footprint on the membrane. The characteristic energy scale
is now chemical in origin, as it is controlled by chemical parameters such as
$\sigma$. It is difficult to give a quantitative scale for this although
typical oil water interfacial
tensions of $\sigma_{ow}\gsim 3\times10^{-2}{\rm Jm}^{-2}$ suggest
$E_o\gsim \sigma_{ow}b^2\approx 180\kT$
with $b\approx 5$nm. This large energy suggests thermodynamically ``strong"
interactions.

\subsubsection{Membrane deformation due to the inclusion}

For $b=10{\rm nm}$, $\sigma=40 dyn/cm$\footnote{The wetting drop is
characterized by a
surface tension, which for highly hydrophobic chains approaches the
oil-water tension
$\sigma=10dyn/cm\ -40dyn/cm=2.5\ -\ 10k_BT/{\rm nm}^2$, and a curvature
radius for the
surface of the drop
$R=1-100{\rm nm}$. These combines to give a pressure $f_b=\sigma/R=2.5\
10^{-2}\ -\ 10
k_BT/{\rm nm}^3$. We arbitrarily choose
$b=R/2$ to give an idea of the scales and define the energy
unit $E_b=f_b b^3=\sigma R^3/8$, so $E_b=0.3 k_BT\ -\ 10^4k_BT$. With
$\sigma_{bio}=40dyn/cm= 10k_BT/{\rm nm}^2$ and $R_{bio}=10{\rm nm}$ we obtain
$f_{b,bio}=1 k_BT/{\rm nm}^3$ and $E_{b,bio}=125 k_BT$. We give only the
results for
$\sigma=40dyn/cm$ and for
$\kappa=20 k_BT$ for compactness. With thses values $k^{-1}=9-90{\rm nm}$.
} the membrane's normal deformation at $r=0$ is
\begin{equation}
u_0(k^{-1}=9{\rm nm})=37{\rm nm}\quad u_0(k^{-1}=30{\rm nm})=65{\rm nm}\quad
u_0(k^{-1}=90{\rm nm})=92{\rm nm}
\end{equation}

for $b=100{\rm nm}$  $\sigma=40 dyn/cm$
\begin{equation}
u_0(k^{-1}=9{\rm nm})=2700{\rm nm}\quad u_0(k^{-1}=30{\rm nm})=15\mu{\rm
m}\quad
u_0(k^{-1}=90{\rm nm})=37\mu{\rm m}
\end{equation}

\subsubsection{Membrane mediated interaction energy between inclusions}

The interaction potential energy has a simple analytical form:
\begin{equation}
\Phi=\frac{1}{2\pi\gamma}\zeta^2K_0(kr)=\frac{E_b^2}{2\pi\kappa}(kb)^2K_0(kr)\beta^2
\end{equation}
with
\begin{equation}
\beta=\frac{2\pi}{(kb)^2}(I_0(kb)-\frac{2}{kb}I_1(kb))
\end{equation}

For $b=10{\rm nm}$ and $\sigma=40 dyn/cm$ the energy scale $E_0$ (simply
the potential
$\Phi$ divided by
$K_0(kr)k_BT$) is
\begin{equation}
E_0(k^{-1}=9{\rm nm})=25\quad E_0(k^{-1}=30{\rm nm})=2\quad
E_0(k^{-1}=90{\rm nm})=0.2
\end{equation}

for $b=100{\rm nm}$  $\sigma=40 dyn/cm$
\begin{equation}
E_0(k^{-1}=9{\rm nm})=10^9\quad E_0(k^{-1}=30{\rm nm})=3\ 10^6\quad
E_0(k^{-1}=90{\rm
nm})=2\ 10^5
\end{equation}

\section{Conclusions}

We show that a model for the interactions between inclusions
bound to fluid membranes can be solved exactly for circularly symmetric
inclusions bound to membranes that are under tension. We argue that proteins
bound to cell membranes can apply no average force, or first moment of
force, to the membrane unless they are also anchored to an external
structure, such as
the cytoskeleton. By this we mean that the total of any and all forces that the
membrabe protein acts to apply downwards against the membrane must exactly
equal the magnitude of similar forces applied upwards. This is a consequence of
Newton's laws of motion.

We proposed idealized models for the origin of these forces due to
either entropic or direct chemical interactions. The
interactions between two circularly symmetric inclusions are {\it repulsive}
and are asymptotically exponentially short ranged with a typical extent
on biological membranes that is of the order of $k^{-1}\approx 30$nm. This
result contrasts with the attraction predicted to appear between similar
inclusions on tensionless membranes. For non circularly symmetric inclusions we
predict an additional algebraically long-ranged attractive contribution.

We discuss how competing attractive and repulsive interactions are known to
sometimes
produce a stripe-like morphology. These phases are reminiscent of the
stripes observed on
the surface of caveloae, which we refer to as ``gnarly buds", and we
discuss how a
similar lateral phase separation could occur in these systems with typical
length scales
comparable to the ranges of the interactions.

We believe that the results discussed here may have wider applications in
understanding
biological phenomena, including lateral phase separation in phospholipid
membranes and
endocytotic budding. Indeed, a detailed theoretical analysis of the
formation of
caveolae invaginations on model phospholipid membranes utilizes many of the
results derived
here~\cite{usPNAS}.

\begin{acknowledgments}
We acknowledge helpful discussions with Robin Ball and George Rowlands
(Warwick), Albert Johner (Strasbourg) and Michael Lisanti (Albert Einstein
College of Medicine). MST
acknowledges the support of the Royal Society (UK) in the form of a
University Research
Fellowship.
\end{acknowledgments}

\section*{Appendix: Uniqueness of the solution for the equilibrium
membrane displacement due to an arbitrary force distribution}

Equations~\ref{usolution} and \ref{ugreens} permit the calculation of the
equilibrium membrane displacement due to an arbitrary force distribution. This
solution is unique, as may be demonstrated as follows.

We may add to our solution any solution of the homogeneous version of equation
(\ref{euler}) without changing the result. Let
$u_1(\r)$ satisfy

\begin{equation}\label{homoeuler}
      \nabla^2\left(\nabla^2 -k^2\right)u_1(\r) = 0
\end{equation}
and the boundary conditions imposed. Multiplying by
$u_1(\r)$ and integrating we have

\begin{equation}
      \int u_1(\r) \nabla^2\left(\nabla^2
      -k^2\right)u_1(\r) d^2\r = 0.
\end{equation}
Integrating by parts and noting that the integrated
terms are zero by virtue of the boundary conditions
we obtain

\begin{equation}
      \int\left[\left(\nabla^2 u_1(\r)\right)^2
      + k^2\left(\nabla u_1(\r)\right)^2 \right] d^2\r=
      0.
\end{equation}
The integrand is everywhere positive so we must have
$\nabla^2u_1(\r)=0$ and $\nabla u_1(\r)=0$
everywhere. Thus the only admissible solution of
the homogeneous equation is $u(\r)$ = constant this
correspond only to a redefinition of the zero of the
displacement and so without loss of generality we take
this constant to be zero. This freedom of choice is a
consequence  of the translational symmetry of the
problem in the direction perpendicular to the
planar membrane.



\newpage

\end{document}